# DESIGN OF A HIGH-POWER COMPACT SRF LINAC FOR INDUSTRIAL APPLICATIONS OF E-BEAM IRRADIATION

J.C.T. Thangaraj* and R.C. Dhuley, Fermilab, Batavia, IL 60510, United States

*Abstract*

Fermilab has developed a novel concept for an industrial electron linac using $Nb_3Sn$ coating technology and conduction cooling. These conduction-cooled linacs can generate electron beam energies up to 10 MeV in continuous-wave operation and reach higher power (>=1 MW) by combing several modules. Compact and light enough to mount on mobile platforms, our machine is anticipated to enable new in-situ environmental remediation applications such as waste-water treatment for urban areas, X-ray medical device sterilization, and innovative pavement applications. We highlight a few aspects of a 1 MW design of such a machine in this paper. A detailed plan is in [1].

## BACKGROUND

Superconducting RF cavities made of $Nb_3Sn$, with cryogenic operation near the temperature of 4 K, exhibit minimal RF wall dissipation (about six orders of magnitude smaller than copper cavities of similar shape and size), allowing their operation at 100% RF duty cycle (continuous wave or CW operation). SRF accelerators can generate very high average power e-beams suitable for high-volume irradiation applications for altering materials' physical, chemical, molecular, and biological properties. These include polymerization, medical and food sterilization, environmental remediation, wastewater treatment, sludge, and biosolids treatment [2]. This paper briefly describes the design of an e-beam accelerator designed to treat high volume (several million gallons per day) of municipal wastewater. The accelerator uses a cryocooled elliptical 650 MHz $Nb_3Sn$ SRF cavity, designed to produce a 10 MeV, 1 MW average power electron beam.

## ACCELERATOR DESIGN

The accelerator layout is divided into three sections (Fig 1): preaccelerator, accelerator (called cryomodule), and the beam delivery system. The preaccelerator comprises a thermionic electron source (gun), an RF injector cavity, and a focusing solenoid magnet. The electron beam exiting the preaccelerator is fed into the accelerator, which energizes the beam to the 10 MeV target energy. The accelerator uses an $Nb_3Sn$ (SRF) cavity operating near 4 K, conduction cooled via cryocoolers. Two fundamental power couplers pierce the vacuum vessel through two ports at 180° to each other to feed RF power into the SRF cavity. The beam exits the accelerator with 10 MeV energy. It then enters the beam delivery system, where it is conditioned using a raster magnet and beam horn for irradiating a stream of wastewater.

___________________
* jtobin@fnal.gov

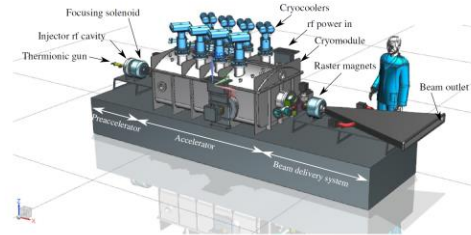

Figure 1: Graphic rendering of an industrial SRF E-beam accelerator components and layout. The overall size is ~4 m long (end-to-end), ~2 m wide, and ~2 m tall.

### Electron Gun

The preaccelerator herein uses a triode RF gun with a gridded thermionic cathode. In this gun, the cathode emits low-energy electrons via thermal emission, which are then shaped into electron bunches using the RF voltage applied to the grid-cathode gap, superimposed on a constant dc voltage. The emitted electrons are then captured and accelerated by the electric field of the RFRF gun. The gun's operating RF amplitude and phase interval are determined for producing a 100 mA average current with a 154 pC electron-beam bunch charge. E-gun parameters are in Fig 2.

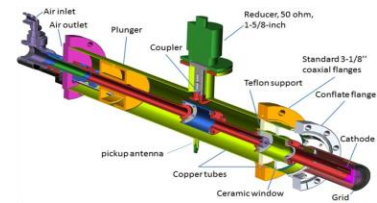

| Parameters | Unit | Value |
|---|---|---|
| Frequency | MHz | 650 |
| Cathode diameter | inch | 0.5 |
| Beam current | mA | 100 |
| Current density | A/cm$^2$ | 2.35 |
| DC bias voltage | kV | 2.6 |
| Output Energy | keV | 3.5 |
| Bunch rms size | Deg | <15 |
| Energy rms size | % | <25 |

Figure 2: RF gun design parameters and components

### Injector cavity

The injector cavity, located immediately downstream of the RF gun, captures the thermionically emitted electrons and accelerates them to ~300 keV energy. The 650 MHz cavity RF design aims to maximize the shunt impedance to get the required accelerating voltage with minimum heat dissipation. Taking copper as the injector cavity material, the voltage gain of ~300 kV would dissipate 11.6 kW of heat, which can be extracted using a forced flow of cooling water around the cavity.

### RF cavity design

We base our design on five-cell, 650 MHz cavities designed to produce the 10 MeV electron beam. The cavity inlet port has a 35 mm diameter, equal to that of the injector cavity outlet. This is much larger than that location's beam spot size of ~12 mm. To match the phase of low-beta electrons entering the cavity, the cavity's first cell has a shorter length than the other four cells. Cells 2–4 have the same length and diameter, while the fifth cell is longer and larger in diameter. The outlet iris of the fifth cell and the downstream beam pipe is also larger in diameter compared to the other four irises. This larger size is chosen to achieve adequate coupling of the fundamental power coupler with the five-cell cavity and out propagation of higher-order modes. The outlet beam pipe has two coupler ports, placed 180° from each outer, for feeding RF power to the cavity.

*Simulation of beam transport through the cavity*

The optimized bunch distribution at the output of the injector cavity is used at the entrance of the five-cell SRF cavity to simulate the bunch acceleration through the five-cell cavity. This assumes negligible bunch distortion between the output of the injector cavity and inlet of the five-cell cavity, facilitated by the focusing solenoid. Note that under this assumption, we have excluded the focusing solenoid from beam transport simulations. Three sets of fields are used in MICHELLE beam transport simulations. The amplitude and phase shift in the five-cell cavity are matched to obtain the 10 MeV beam at the outlet of the accelerator.

Using up to 100 000 particles and neglecting particle loss to cavity walls, the MICHELLE simulations confirmed that the five-cell cavity produces uniform acceleration in each cell, starting from 0.3 MeV at the injector outlet and ending at 10 MeV at the five-cell cavity outlet. Finally, the beam characteristics and particle distributions at the exit of the five-cell cavity are summarized in the plots in Fig. 3.

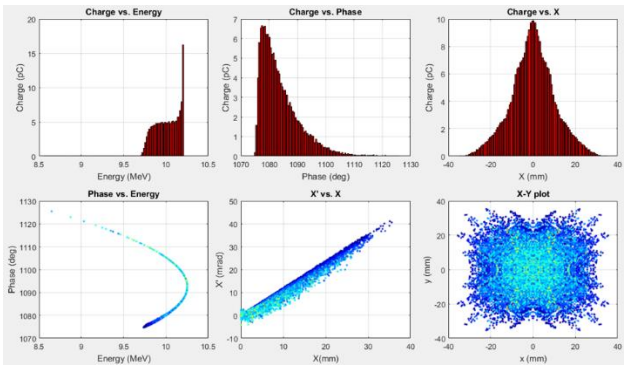

Figure 3: Beam characteristics and particle distributions at the exit of the 5-cell cavity. Top plots: charge distribution vs. energy and phase, current vs. radius; bottom plots: phase vs. energy, x´-x phase-space distribution, beam spot size.

*Design of Fundamental Power Coupler (FPC)*

The RF power is fed to the five-cell SRF cavity using two couplers, each sustaining 500 kW CW power with ~10% reflection. This design uses a ceramic window (disc) to separate the air side of the coupler that receives power from an RF source, from the vacuum side that delivers power to the cavity. The disk is made of alumina with a loss tangent of $10^{-4}$ or less and is brazed with the outer and inner copper bushings. The outer copper bushing is brazed with a stainless-steel ring connecting to the coupler's outer conductor.

To reduce static and dynamic cryogenic loading to 4 K, the coupler uses a copper electromagnetic shield (EMS) heat sunk only at ~50 K and no physical contact with the cavity. The EMS screens the stainless-steel outer conductor, aluminum gasket, and stainless flange from the electromagnetic field, thereby reducing Ohmic losses in the outer conductor, gasket, and flange. All RF losses are primarily concentrated in the EMS and are intercepted by the 50 K thermal intercept. The EMS also includes an iris to reduce thermal radiation from room temperature ceramic and protects the ceramic window from charged particles from the cavity. The inner conductor of the coupler is a hollow copper channel cooled with a forced flow of water. Thermal analysis is conducted to estimate cryo loading (Ohmic losses) in the coupler under steady operation with 500 kW forward propagation. The antenna tip is expected to operate at ~312 K with water cooling.

*Cryomodule design*

The cryomodule includes a vacuum vessel, a 650 MHz $Nb_3Sn$ cavity, eight 4 K cryocoolers, and single-layer thermal and magnetic shields. The cavity is conduction-cooled to the cryocoolers following the procedure described in [3-6]. The thermal shield is connected to the 50 K cooling stage of the cryocoolers using a set of thermal straps [7]. Two types of cryocoolers are selected: six Cryomech PT420 offering a cooling capacity of 2.0 W at 4.2 K and two additional Cryomech PT425 with a higher cooling capacity of 2.5 W at 4.2 K located above the RF couplers. Strong magnetic fields can impair the intrinsic quality factor of the cavity, thereby reducing the attainable accelerating gradient for a given cryocooling capacity. A magnetic shield is provided to limit the total magnetic field on the surface of the SRF cavity to <10 mG. The magnetic shield is operated at room temperature to avoid additional cryogenic loading of the cryocoolers. The cold mass and the magnetic shield are all enclosed in a 1.95-m-long vacuum vessel. The total mass of the fully assembled cryomodule is estimated to be 1750 kg.

*Cavity cryogenic thermal management*

The cavity is divided into two sections to simplify the heat load estimation: (i) the main body comprising the five elliptical cells and the inlet beam tube, and (ii) the outlet side made of the two coupler ports and outlet beam pipe. The static heat leak contributions of thermal radiation from the thermal shield and via beam pipes, thermal conduction via cavity supports, beam pipes, and coupler ports are considered. The dynamic loading comprises beam loss (assumed 1 W), cavity RF heating (calculated using BCS resistance of $Nb_3Sn$ [8] plus 10 nΩ residual), and coupler loading (taken to be 3 W per coupler as shown for a coupler developed at BNL for similar power level [9]). Combining the static and dynamic heat leaks, the cavity is loaded by 19.5 W. The cavity body and inlet side experience 14 W of heat load, which can be extracted using six Cryomech PT420 cryocoolers. The cavity outlet side sees 6.5 W of cryo loading, which is extractable using two units of Cryomech PT425 cryo-coolers. 5N purity aluminum links mechanically couple the five-cell niobium cavity with the cryocoolers. The link components are cut out of commercial aluminum sheets, bent into final shapes, and carry holes for bolting to the cryocoolers and the cavity.

*Thermal and magnetic shields*

The thermal shield performance is evaluated using a heat transfer analysis in COMSOL Multiphysics. The major heat loads on the thermal shield are thermal radiation, interception of FPC, and interception of inlet and outlet beam pipes totaling 138 W. The thermal shield is coupled to the 50-K stages of the eight cryocoolers, providing sufficient cooling capacity to hold the thermal shield at a spatial uniform temperature of ~43 K. Also, COMSOL simulation of magnetic field attenuation using a single-layer 2 mm thick mu-metal shield around the cavity demonstrated total magnetic fields of ~7 mG on the surface of the SRF cavity, which is below the target field of <10 mG.

*Vacuum vessel*

The cryomodule vacuum vessel is made of 316L stainless steel and has a detachable bottom tub and a top lid. The vacuum seal is established using an o-ring along the periphery, pressed using bolted connections. The vacuum vessel walls are 5/16" thick, and the structure is reinforced on the outside by 3/8"-to 1/2"-thick stiffeners that prevent buckling under external pressure. The total weight of the cryostat vacuum vessel is approximately 462 kg, *i.e.*, 168 kg for the lid and 293 kg for the tub. The vacuum vessel is designed following ASME BPVC VIII.2 Alternative Design Rules, satisfying the criteria for protection against plastic collapse, local failure, and buckling under external pressure [10].

## SUMMARY

We have briefly presented the beam dynamics, RF, thermal, and engineering design of a 10 MeV, 1 MW average power e-beam accelerator driven by a room temperature pre-accelerator and a conduction-cooled SRF accelerator cryomodule. We encourage the interested reader for a more detailed design of the accelerator and modeling of its capital and operating cost to refer to [1].

## ACKNOWLEDGMENTS

This work was supported by Fermi Research Alliance, LLC under Contract No. DE-AC02-07CH11359 with the U.S. Department of Energy, Office of Science, Office of High Energy Physics. Research funded by DOE HEP Accelerator Stewardship. Thanks to Tom Kroc, Mauricio Suarez, Christopher Edwards, and Yichen Ji for their review and comments.